# Integrative omics framework for characterization of coral reef ecosystems from the *Tara Pacific* expedition

Caroline Belser[1*], Julie Poulain[1*], Karine Labadie[2], Frederick Gavory[1], Adriana Alberti[1,3], Julie Guy[2], Quentin Carradec[1], Corinne Cruaud[2], Corinne Da Silva[1], Stefan Engelen[1], Paul Mielle[4], Aude Perdereau[2], Gaelle Samson[2], Shahinaz Gaz[2], Genoscope Technical Team[2], Christian R Voolstra[5], Pierre E. Galand[6], J. Michel Flores[7], Benjamin CC Hume[5], Gabriela Perna[5], Maren Ziegler[8], Hans-Joachim Ruscheweyh[9], Emilie Boissin[10], Sarah Romac[11], Guillaume Bourdin[12], Guillaume Iwankow[10], Clémentine Moulin[13], David A. Paz García[14], Tara Pacific Consortium Coordinators, Claude Scarpelli[2], E'Krame Jacoby[2], Pedro H. Oliveira[2], Jean-Marc Aury[1], Denis Allemand[15], Serge Planes[10], Patrick Wincker[1]

* These authors contributed equally

corresponding author: Caroline Belser (cbelser@genoscope.cns.fr)

**Affiliations**

[1]Génomique Métabolique, Genoscope, Institut François Jacob, CEA, CNRS, Univ Evry, Université Paris-Saclay, Evry, France

[2]Genoscope, Institut François Jacob, Commissariat à l'Energie Atomique (CEA), Université Paris-Saclay, 2 Rue Gaston Crémieux, 91057 Evry, France

[3]Current address: Université Paris-Saclay, CEA, CNRS, Institute for Integrative Biology of the Cell (I2BC), 91198, Gif-sur-Yvette, France.

[4]Centre National de Recherche en Génomique Humaine (CNRGH), Institut de Biologie François Jacob, CEA, Université Paris-Saclay, 91000 Evry, France.

[5]Department of Biology, University of Konstanz, 78457 Konstanz, Germany

[6]Sorbonne Université, CNRS, Laboratoire d'Ecogéochimie des Environnements Benthiques, LECOB, Observatoire Océanologique, 66650, Banyuls/Mer, France

[7]Weizmann Institute of Science, Dept. Earth and Planetary Science, Rehovot, Israel

[8]Department of Animal Ecology & Systematics, Justus Liebig University Giessen, 35392 Giessen, Germany

[9]Department of Biology, Institute of Microbiology and Swiss Institute of Bioinformatics, ETH Zürich, Zürich 8093, Switzerland

[10]PSL Research University: EPHE-UPVD-CNRS, USR 3278 CRIOBE, Laboratoire d'Excellence CORAIL, Université de Perpignan, 52 Avenue Paul Alduy, 66860 Perpignan Cedex, France

[11]Sorbonne Université, CNRS, Station Biologique de Roscoff, AD2M, UMR 7144, ECOMAP, Roscoff, France

[12]School of Marine Sciences, University of Maine, USA

[13]Fondation Tara Océan, Base Tara, 8 rue de Prague, 75 012 Paris, France

[14]CONACYT-Centro de Investigaciones Biológicas del Noroeste. La Paz, Baja California Sur, 23096, México.

[15]Centre Scientifique de Monaco, 8 Quai Antoine Ier, MC-98000, Principality of Monaco

[Δ] The *Genoscope Technical Team* is listed at the end of this manuscript

[Δ] The *Tara Pacific Consortium Coordinators* and their affiliations are listed at the end of this manuscript

## Abstract

Coral reef science is a fast-growing field propelled by the need to better understand coral health and resilience to devise strategies to slow reef loss resulting from environmental stresses. Key to coral resilience are the symbiotic interactions established within a complex holobiont, *i.e*. the multipartite assemblages comprising the host coral organism, endosymbiotic dinoflagellates, bacteria, archaea, fungi, and viruses. Tara Pacific is an ambitious project built upon the experience of previous Tara Oceans expeditions, and leveraging state-of-the-art sequencing technologies and analyses to dissect the biodiversity and biocomplexity of the coral holobiont screened across most archipelagos spread throughout the entire Pacific Ocean. Here we detail the Tara Pacific workflow for multi-omics data generation, from sample handling to nucleotide sequence data generation and deposition. This unique multidimensional framework also includes a large amount of concomitant metadata collected side-by-side that provide new assessments of coral reef biodiversity including micro-biodiversity and shape future investigations of coral reef dynamics and their fate in the Anthropocene.

## Background & Summary

Anthropogenic climate change, including ocean warming, acidification, and deoxygenation profoundly alter the ocean's environmental conditions, leading to structural transformation of marine ecosystems. Coral reefs are particularly vulnerable because their structural integrity is particularly sensitive to these factors and recent works have demonstrated significant losses of coral reefs [1-3]. Their biology and ecology are dependent on the presence of healthy coral holobionts, a diverse and fragile consortium of microbial communities including the coral host, endosymbiotic microalgae, bacteria, archaea, fungi and viruses whose dynamics are altered during environmental disturbance [4]. Dysbiosis in the coral holobiont microbiota may also have substantial consequences on the diversity of reef-associated animal communities (e.g., crustaceans, molluscs, fishes), which depend on corals for food, shelter, and reproduction [5,6]. Hence, methods that jointly leverage multi-omics data to disentangle the biocomplexity of corals and their associated microbial symbionts (holobiont-omics) will provide key insight into resilience, acclimatization, and environmental adaptation of corals and coral reefs [7].

In light of the anthropogenic threats to coral reef systems, the Tara Pacific expedition (2016–2018) was launched as a cooperative international effort to overview the omics complexity of the coral holobiont and its ecosystem across the entire Pacific Ocean with a screening of most archipelagos [8]. Inspired by previous Tara Oceans expeditions [9,10], Tara Pacific undertook the first pan-ecosystemic study of coral reef diversity throughout the Pacific Ocean, drawing an east–west transect from Panama to Papua New Guinea and a south–north transect from Australia to Japan, sampling corals throughout 32 island systems [8]. The project targeted two scleractinian corals (*Pocillopora meandrina* and *Porites lobata*), one hydrocoral (*Millepora platyphylla*), and two species of reef fish (*Acanthurus triostegus* and *Zanclus cornutus*), chosen for their presence throughout the Pacific. Sampling also included near-island, reef surface and coral surrounding water samples, as well as open ocean water and air samples collected in-between islands [11]. The global sampling strategy and the contextual metadata of the Tara Pacific expeditions are presented in papers previously published [8,12].

Such large-scale biodiversity monitoring programs, particularly those focusing on in-depth microbial composition assessments from environmental DNA, are crucial to assess trends in spatial structure and temporal turnover of environmentally perturbed communities since microbial communities are more reactive and may give early biodiversity indicators of community change. In this regard, high-throughput DNA and RNA sequencing approaches (e.g., metabarcoding, metagenomics, metatranscriptomics) hold considerable promise for a fundamental understanding of the biodiversity and for monitoring its global rates of biodiversity change in the coral holobiont and tropical reefs at large (Fig. **1**). For example, metabarcoding (marker gene) strategies will greatly enhance our understanding of the biodiversity (microbiome) associated with and surrounding the coral holobiont as well as reef fish. Further, dual RNA-Seq for example, allows characterizing the relationships among endosymbiotic microalgae and the coral animal, whereas analyses of the metagenome and metatranscriptome of the water column surrounding the coral can provide insight into how regional settings and biological communities influence and structure microbiome dynamics. Here we present detailed protocols used in the Tara Pacific expedition, from sample handling to nucleotide sequence data deposition. This multiscale framework will hopefully help to direct future research foci and lead to a better understanding of the critical biotic interactions that underlie reef health.

# Methods

The workflow presented here consists of a set of optimized, automatable, and cost-competitive protocols adapted to each of the multiple components of the reef (*e.g.*, coral, fish, plankton). As marine biomonitoring increasingly moves towards an ecosystem-based approach, the richness of this omics data can be leveraged as an indicator of how biotic networks and coral reefs are impacted by anthropogenic activities. A global overview of the protocols and corresponding omics analysis is presented in Table **1** and detailed in the following sections. All the abbreviations used in this study can be found in Table **S1**.

## 1. Handling of biological samples

Biological samples were transferred from 12 ports of call (locations allowing dry ice cargo in the Pacific) to the French National Sequencing Center (CEA-Genoscope, Evry). Appropriate and uniform refrigeration was assured from the boat to the final lab storage rooms using dry ice. Upon arrival, samples were readily identified by scanning or reading their unique identifier label (ID barcode for sample tracking). Each unique barcode was generated upstream of the sampling stage, and allowed precise linkages between the samples to their corresponding rich and varied metadata (*e.g.*, sampling date and location, taxa, etc) [Data citations #1]. In addition, samples were registered in the ENA BioSample database [Data citations #2].

All manipulations for each sample were recorded and will be available through an in-house-developed Next Generation Laboratory Information Management System (called NGL for Next Generation LIMS). The architecture of NGL is described in detail in **Technical Validation Section 1**. Samples intended for sequencing were stored at the appropriate temperature (either room temperature, +4 °C, -20 °C, or −80 °C), whereas the remainder were repackaged and forwarded to the different laboratories responsible for different analysis (metabolomic, aging, telomere, stress markers, among others [8]).

## 2. Nucleic acid extraction

Different nucleic acid extraction strategies were developed, depending on sample type and sequencing strategy envisaged. An optimized and benchmarked two-step process was put in place for plankton (size fractions S0.2-3µm, S3-20µm, S>20µm and S>300µm), coral (*Pocillopora meandrina*, *Porites lobata*, *Millepora platyphylla*), coral diversity (CDIV, sampling of all the coral diversity at each island), and fish samples (mucus and gut of *Acanthurus triostegus* and *Zanclus cornutus*), whereas virus, aerosol and sediment samples were processed using extraction protocols implemented respectively by independent work in the Sullivan, Flores and Voolstra laboratories.

### *2.1. Cell disruption*

Two strategies were employed for cell disruption: *i*) cryogenic grinding (cryogrinding) in the case of planktonic organisms (*i.e*. water filter samples) and *ii*) matrix/bead beating in the case of coral and fish samples.

#### *2.1.1. Cryogenic grinding*

Cryopreserved membrane filters were ground in order to disrupt cells, even the most resistant such as silica-based ones. Each membrane was accommodated into a grinding vial with 1 ml RA1 lysis buffer (Macherey-Nagel, Düren, Germany) and 1% β-mercaptoethanol (Sigma, St Louis, MO, USA) and subjected to the cryogenic freezer mill (SPEX Sample Prep, Metuchen, New Jersey, USA) with the following grinding program: 2 min of pre-cooling time, first grinding cycle at 10 knocks/s for 1 min, 1 min of cooling time, and a final grinding cycle at 10 knocks/s for 1 min. The cryoground powder was then subjected to nucleic acid extraction (Section **2.2.1**).

#### *2.1.2. Bead beating*

Coral and CDIV samples were respectively collected in 15 ml and 2 ml of Lysing Matrix A beads (MP Biomedicals, Santa Ana, CA, USA), whereas fish samples were collected in 2 ml tubes. All were preserved in the presence of DNA/RNA Shield buffer (Zymo Research, Irvine, CA, USA), respectively 1.5 ml for 2 ml tube and 10 ml for 15 ml tube and stored at -20 °C. Prior to extraction, samples were thawed for 30 min at room temperature. Fish samples were transferred using forceps into a 2 ml ZR Bashing Bead Lysis tube. DNA/RNA shield buffer was transferred by pipetting. The initial sample tube was rinsed with 500 µl of DNA/RNA shield and transferred to the ZR Bashing bead lysis tube. Coral cells, CDIV, and fish samples were then disrupted by the simultaneous multidirectional striking of the lysis Matrix A beads or ZR Bashing bead using a high-speed homogenizer FastPrep-24 5G Instrument (MP Biomedicals, Santa Ana, CA, USA) under the following conditions: speed: 6.0 m/s, time: 30 s, pause time: 60 s, cycles: 3. The homogenized sample was left to stand for 15 min at room temperature (allowing foam to disappear and large particles to settle), and then split as follows: ten aliquots of 500 µl for coral, two aliquots of 200 µl for CDIV, and two aliquots of 500 µl for fish samples. Aliquots were stored at -20 °C until further nucleic acid extractions and purification. To test for potential contamination occurring during the extraction process, we included a negative control (NC) where 10 ml or 1 ml of DNA/RNA shield were respectively transferred to 15 ml Lysing Matrix A beads or 2 ml ZR Bashing Bead Lysis tubes and ground as described previously. Purified DNA from these NC extractions were checked by metabarcoding.

### *2.2. DNA/RNA extraction*

#### *2.2.1. DNA/RNA extraction from planktonic organisms*

The cryoground powder was subjected to nucleic acid purification with the NucleoSpin RNA kit (Macherey-Nagel, Düren, Germany) combined with the DNA Elution buffer kit (Macherey-Nagel, Düren, Germany). Briefly, the cryoground powder was resuspended in 2 ml RA1 lysis buffer with 1 % β-mercaptoethanol, transferred to a large capacity NucleoSpin Filter from the RNA Midi kit, and centrifuged for 10 min at 1,500 *g*. After further addition of 1 ml RA1 lysis buffer with 1 % β-mercaptoethanol, the filter was centrifuged 3 min at 1,500 *g*. The eluate was transferred to a new tube with addition of 1 volume of 70% ethanol, and the mixture was loaded to a NucleoSpin RNA Mini spin column and washed twice with DNA washing solution. DNA was eluted three times, each with 100 µl DNA elution buffer, and stored at −20 °C. RNA purification was continued on the previous NucleoSpin RNA Mini spin column by digesting residual DNA with 10 µl rDNase and 90 µl reaction buffer. After 15 min of incubation at room temperature, the column was washed with RA2 and RA3 buffers. RNA was eluted in 60 µl RNase-free water and stored in sterile microtubes at −80 °C. The quantity and quality of the extracted RNA was assessed by fluorometric quantitation on a Qubit 2.0 Fluorometer using the Qubit RNA HS Assay (Thermo Fisher Scientific, Waltham, MA, USA). The quality of total RNA was checked by capillary electrophoresis on an Agilent Bioanalyzer using the RNA 6,000 Pico LabChip kit (Agilent Technologies, Santa Clara, CA, USA). To maximally reduce the risk of presence of residual genomic DNA (gDNA), leading to misinterpretation of RNA sequencing data, an extra DNase treatment was applied on the total RNA extracted, beyond the in-column DNase treatment already included in the extraction procedure. Total RNA samples were further processed as follows: a quantity of approximately 5 µg was treated with the Turbo DNA-free kit (Thermo Fisher Scientific, Waltham, MA, USA), according to the manufacturer's protocol. After two rounds of incubation at 37 °C for 30 min, the reaction mixture was purified with the RNA Clean and Concentrator-5 kit (Zymo Research, Irvine, CA, USA) following the procedure described for retention of >17 nt RNA fragments. RNA was eluted in 9-15 µl nuclease-free water by two elution steps in order to maximize recovery. After purification, DNA and RNA were submitted to quality control assessment as described in Sections **3.1** and **3.2**. To evaluate a potential contamination occurring during the extraction process, a NC consisting of a blank filter was submitted to the same extraction procedure described above.

### 2.2.2 DNA/RNA extraction from coral samples

To apply diverse sequencing strategies (i.e., metagenomic, metatranscriptomic, and metabarcoding) for the coral host, endosymbiont microalgae, coral microbiome, and virome, we applied two versions of the nucleic acid purification protocol after the bead-beating step.

These versions (detailed below) relied upon the commercial Quick-DNA/RNA Kit (Zymo Research, Irvine, CA, USA), but one of them was supplemented by an enzymatic digestion step in order to achieve an optimal lysis of the bacterial and archaeal components of the microbiome.

#### *2.2.2.1 RNA extraction for dual compartment (coral/endosymbiont) transcriptome profiling*

This first version of the protocol was designed to obtain high-quality RNA of coral host and endosymbiont microalgae for a dual-transcriptomic strategy. Essentially, one aliquot of 500 µl of homogenized suspension was thawed at room temperature and transferred to a 2 ml tube in the presence of Digestion Buffer Proteinase K (50 µl) and Proteinase K (25 µl, 20 mg / ml). After an incubation of 30 min at 55 °C, the mixture was transferred to a 5 ml tube in the presence of DNA/RNA Lysis Buffer (1.5 ml). The mixture was vortexed and let stand 15 min at room temperature. 700 µl of the mixture was then transferred into a Spin-Away Filter in a collection tube and centrifuged for 30 s at 14,000 *g*. The flow-through was collected in a 5 ml tube for RNA purification, while the Spin-Away Filter1 was kept for DNA purification. This step was repeated until the passage of the entire volume. 2 ml of 100 % ethanol was added to the flow-through and mixed manually. Precipitated RNA solution (700 µl) was added into a Zymo Spin III CG placed in a collection tube and centrifuged for 30 s at 14,000 *g*. The flow-through was discarded and the step repeated until passage of the entire volume. DNA/RNA Prep Buffer (400µl) was added to each column. The Spin-Away Filter1 and Zymo Spin III CG were centrifuged for 30 s at 14,000 *g* and the flow-through was discarded. Columns were loaded with 700 µl of DNA/RNA Wash Buffer and centrifuged for 30 s at 14,000 *g*. They were loaded again with DNA/RNA Wash Buffer (400 µl) and centrifuged for 2 min at 14,000 *g*. RNA elution was finally achieved by adding DNAse/RNAse-free water (100 µl). The tube was incubated 5 min at room temperature and centrifuged for 30 s at 14,000 *g*. Purified RNA was stored at -80 °C until further use.

### *2.2.2.2 DNA extraction for metagenomic and metabarcoding assays (coral/endosymbiont/associated microbiome)*

This second version of the protocol is similar to the one described in **2.2.2.1** but supplemented by an additional enzymatic digestion step with the goal of achieving an optimal lysis of the bacterial and archaeal components of the microbiome. Essentially, 500 µl of homogenized suspension was thawed at room temperature and then transferred to a 2 ml tube in the presence of 50 µl of lysozyme (50 µl 10 mg/ml), 3 µl of mutanolysine (50 KU/ml) and 3 µl of lysostaphine (4 KU/ml) [13,14]. After 1 h incubation at 37 °C in a thermoblock, 50 µl of Digestion Buffer Proteinase K and 25 µl of Proteinase K (20 mg/ml) were added, and a second incubation was performed for 30 min at 55 °C. The mixture was finally transferred to 5 ml tubes and the protocol was continued as described in the Section **2.2.2.1**. Purified DNA was stored at -20 °C.

### *2.2.3. DNA extraction from CDIV samples*

Since CDIV samples were solely intended for metabarcoding, and because their number was relatively high, we opted for a 96 deep-well-based DNA extraction protocol containing using the Quick-DNA 96 Plus kit (Zymo Research, Irvine, CA, USA). The same additional enzymatic digestion step described in the Section **2.2.2.2** was applied to obtain an optimal lysis of prokaryotic cells. Briefly, 200 µl of homogenized suspension from 95 CDIV samples and 1 aliquot from the grinding NC were thawed at room temperature and transferred to a 96 deep-well plate. DNA/RNA shield (200 µl) was added to the last well, which served as an extraction NC. Lysozyme (20 µl, 10 mg / ml), mutanolysine (1.5 µl, 50 KU / ml) and lysostaphine (1.5 µl, 4 KU / ml) were added to each well to achieve microbial lysis [13,14]. The plate was sealed using an aluminum pad and incubated for 1 h at 37 °C. The plate was briefly spun and Proteinase K (10 µl, 20 mg / ml) was added to each well. The sealed plate was incubated a second time for 30 min at 55 °C. It was then briefly spun, and in case debris was still present, it was centrifuged for 5 min at 1,000 *g*, followed by transferring 233 µl to a new deep-well plate.

Genomic Binding Buffer (233 µl) was added to each well and gently mixed by pipetting. The entire volume of each well was transferred to a Zymo-Spin I-96 XL plate placed into a collection plate. DNA Pre Wash Buffer (200 µl) was added, the plate was centrifuged 5 min at 3,500 *g*, and the flow-through was removed. Next, gDNA wash buffer (500 µl) was added, the plate was centrifuged 5 min at 3,500 *g*, and the flow-through discarded. This last step was repeated with 200 µl of gDNA wash buffer, and then the Zymo-Spin I-96 XL plate was placed into an elution plate. Wells were loaded with DNAse/RNase-free water (25 µl) prewarmed at 50 °C. After 5 min, the plate was centrifuged for 5 min at 3,500 *g*. This last step was repeated once and the DNA eluates were kept in the plate and stored at -20 °C.

### 2.2.4. DNA extraction from fish gut and mucus samples

One aliquot of 500 µl of homogenized suspension was thawed at room temperature and transferred to a 2 ml tube in the presence of Digestion Buffer Proteinase K (50 µl) and Proteinase K (25 µl, 20 mg/ml). After an incubation of 30 min at 55 °C, DNA was extracted using the Quick-DNA/RNA Miniprep plus Kit (Zymo Research, Irvine, CA, USA) described above. Briefly, the mixture was transferred into a 5 ml tube in the presence of DNA/RNA Lysis Buffer (1.5 ml), then vortexed and let sit for 15 min at room temperature. 700 µl of the mixture was then transferred into a Spin-Away Filter1 in a collection tube and centrifuged for 30 s at 14,000 *g*. The flow-through was collected in a 5 ml tube for RNA purification and the Spin-Away Filter1 was kept for DNA purification. This step was repeated until the passage of the entire volume. 100 % ethanol (2 ml) was added to the eluate and manually inverted. The precipitated RNA solution (700 µl) was added into a Zymo spin III CG placed in a collection tube and centrifuged for 30 s at 14,000 *g*. The flow-through was discarded and the step repeated until the passage of the entire volume. DNA/RNA Prep Buffer (400 µl) was added to each column. The Spin-Away Filter1 and Zymo spin III CG were centrifuged (30 s, 14,000 *g*) and the flow-through was discarded. Columns were loaded with DNA/RNA Wash Buffer (700 µl) and centrifuged (30 s, 14,000 *g*). They were loaded again with DNA/RNA Wash Buffer (400 µl) and centrifuged for 2 min at 14,000 *g*. Elution of DNA and RNA was achieved by adding

DNAse/RNAse-free water (100 µl) into the columns, followed by incubation for 5 min at room temperature and centrifuging for 30 s at 14,000 *g*. DNA and RNA were stored respectively at -20 °C and at -80 °C.

### 2.2.5. DNA extraction from viral particles

This section describes the recovery of viruses from seawater using iron-based flocculation and large-pore-size (0.22 µm) filtration, followed by resuspension of the virus-containing precipitate in a pH 6 buffer. This Fe-based virus flocculation, filtration and resuspension method (FFR) is efficient (>90 % recovery), reliable, inexpensive, and suited for marine viral ecology and genomics research [15]. Briefly, $FeCl_3$ precipitation was used to concentrate viruses from 20-60 l of 0.22 µm filtered seawater, which were then resuspended in ascorbate buffer (0.125 M Tris-base, 0.1 M sodium EDTA dehydrate, 0.2 M $MgCl_2{\cdot}6H_2O$, 0.2 M ascorbate). Following resuspension, recovered viruses were treated with DNase I to remove free DNA [16] 0.1 M EDTA and 0.1 M EGTA to halt DNase activity, and further concentrated to <1 ml using an Amicon 100 kDa filter (Sigma-Aldrich, St. Louis, MO, USA). DNA was extracted using the Wizard Prep DNA Purification system (Promega, Madison, WI, USA). All detailed protocols are listed by name and available at https://www.protocols.io/groups/sullivan-lab. DNA extracted in the Center of Microbiome Science of the Ohio State University, was then sent to Genoscope and submitted to quality control assessment as described in Section **3**.

### 2.2.6. DNA extraction from aerosol samples

DNA was extracted from air filters using the DNeasy PowerWater Kit (Qiagen, Hilden, Germany). Briefly, filters were defrosted and placed into 5 ml bead tubes to which pre-heated (60 °C) PW1 lysis buffer (1 ml) was added. After 5 min of horizontal vortexing, samples were centrifuged (1 min, 4,000 *g*), supernatants were transferred and centrifuged (1 min, 13,000 *g*, 4 °C). Next, the supernatants were transferred into a clean microcentrifuge tube, and additional IRS buffer (200 µl) was added. Samples were mixed and incubated for 5 min on ice, after which they were centrifuged (1 min, 13,000 *g*, 4 °C). The supernatant was transferred into a clean tube and PW3 buffer (650 µl) was added and mixed by pipetting. The supernatant (650 µl) was transferred into a MB spin column placed in a microcentrifuge tube and centrifuged (1 min, 13,000 *g*, 4 °C). Subsequently, the flow-through was discarded and additional 650 µl of supernatant were loaded on the spin column to be centrifuged in the same conditions. This step was repeated until all supernatant was applied on the column (between 2 to 3 times). Next, the spin column filter was transferred into a clean microcentrifuge tube, and pre-shacked PW4 buffer (650 µl) was added to the spin column for another centrifugation step (1 min, 13,000 *g*, 4 °C). The spin column was placed on a clean collection tube and ethanol (650 µl) was added, followed by centrifugation (1 min, 13,000 *g*, 4 °C). After disposal of the flow-through, the empty spin columns were centrifuged (2 min, 13,000 *g*, 4 °C) for disposal of remaining ethanol. The dry spin column was placed on a clean collection tube and EB solution (75 µl) was added to the center of the white filter membrane. The tubes were incubated at room temperature for 1 min after which they were centrifuged (1 min, 13,000 *g*, 4 °C). The elution step was repeated by reintroducing the flow-through eluted DNA solution on the spin column membrane. In each extraction batch, one blank filter was extracted. We also included NC filters with no air sampled on them.

### *2.2.7. DNA extraction from sediment samples*

Sediment samples were first handled by the Reef Genomics Lab at the Red Sea Research Center of the King Abdullah University of Science and Technology. DNA was extracted using the Qiagen DNeasy Plant Mini Kit. Briefly, samples were defrosted, the DNA/RNA Shield buffer was decanted or ethanol removed, and samples air dried for 5 min in order to remove residual ethanol. Samples were then vortexed and for each sample 0.25 g of sediment was transferred into a 1.5 ml Eppendorf tube and AP1 lysis buffer (750 µl) added. After brief vortexing, samples were mixed on a rotating wheel for 30 min. Supernatant (400 µl) was transferred to a new microtube, RNase A (4 µl) was added, samples were vortexed, and incubated for 10 min at 65 °C with tube inversion every 3 min. DNA extractions were then performed according to the manufacturer's instructions, with a final elution volume of 100 µl.

DNA concentrations were quantified on a Qubit 2.0 Fluorometer with the Qubit dsDNA High Sensitivity Assay Kit (ThermoFisher Scientific, Waltham, MA, USA) and DNA samples were sent to Genoscope, France.

## 3. Quality control assessment of DNA/RNA samples

### *3.1. DNA quantification*

DNA was quantified by fluorometry using a Qubit 2.0 Fluorometer instrument with the Qubit dsDNA BR (Broad range) and HS (High sensitivity) Assays (ThermoFisher Scientific, Waltham, MA, USA). Given the very low biomass of aerosol and virus samples, DNA concentration was evaluated by a Qubit spectrophotometer with a DeNovix dsDNA High Sensitivity kit (Denovix, Wilmington, DE, USA).

### *3.2. RNA quantification and qualification*

Quantity and quality of extracted RNA were assessed on a Qubit 2.0 Fluorometer using a Qubit RNA HS Assay kit. The quality of total RNA was checked by capillary electrophoresis on an Agilent Bioanalyzer using the RNA 6,000 Pico LabChip kit (Agilent Technologies, Santa Clara, CA, USA).

## *4. Library preparation for metagenomic samples*

### *4.1. Library preparation for gDNA from size-fractionated filters, coral, and fish gut*

Library preparation protocols were constructed to generate narrow-sized libraries around 300-800 bp. The library preparation protocol was chosen based on the DNA extraction yield as described in Fig. **2**.

#### *4.1.1. Library preparation for DNA quantities > 500 ng*

For samples containing more than 500 ng of total DNA, an aliquot of 250 ng was first sheared to target a mean size of 380 bp using a Covaris E210 instrument (Covaris Inc., Woburn, MA,

USA). Size profiles of sheared materials were visualized on an Agilent Bioanalyzer DNA High Sensitivity chip. The resulting fragmented DNA was end-repaired, A-tailed at the 3'end, and ligated to Illumina compatible adaptors using NEBNext DNA Modules (New England Biolabs, MA, USA) and NextFlex DNA barcodes (BiOO Scientific Corporation, Austin, TX, USA) with our in-house-developed 'on beads' protocol [17]. A liquid handler, the Biomek FX Laboratory Automation Workstation (Beckman Coulter Genomics, Danvers, MA, USA), was used to perform up to 96 reactions in parallel. After two consecutive 1x AMPure XP clean-ups, the ligated products were amplified using Kapa Hifi HotStart NGS library Amplification kit (Kapa Biosystems, Wilmington, MA, USA), followed by 0.6x AMPure XP purification.

#### *4.1.2. Library preparation for DNA quantities of 250-500 ng*

When samples contained between 250 to 500 ng of total DNA, an aliquot of 50–100 ng was used in the shearing step. Before 2018 we used the NebNext protocol described above but in a manual mode. In 2018, a new protocol adapted to low input DNA was implemented using a liquid handler, the Biomek FX Laboratory Automation Workstation (Beckman Coulter Genomics). Fragments were end-repaired, 3'-adenylated and NEXTflex DNA barcoded adaptors (BiOO Scientific Corporation, Austin, TX, USA) were added by using NEBNext Ultra II DNA Library prep kit for Illumina (New England Biolabs, Ipswich, MA, USA). After two consecutive 1x AMPure clean ups, the ligated products were PCR-amplified with the NEBNext® Ultra II Q5 Master Mix included in the kit, followed by 0.8x AMPure XP purification.

#### *4.1.3. Library preparation for DNA quantities <250 ng*

When the extraction yielded low DNA quantities, 10-50 ng of total DNA were sonicated and the NEBNext Ultra II DNA Library prep kit for Illumina was manually applied. Fragments were end-repaired, 3'-adenylated and NEXTflex DNA barcoded adaptors were added by using the NEBNext Ultra II DNA Library prep kit for Illumina. After two consecutive 1x AMPure clean ups, the ligated products were PCR-amplified with the NEBNext Ultra II Q5 Master Mix included in the kit, followed by 0.8x AMPure XP purification.

### *4.2. Library preparation from viral samples*

Recent research has highlighted the existence and potential high diversity of ssDNA viruses in marine ecosystems, as well as the lack of information about their relative abundance due to technical limits in their analysis [18,19]. Advances in library preparation methods now allow the simultaneous recovery of dsDNA and ssDNA for sequencing analysis. The library preparation protocol chosen for the processing of viral samples benefited from benchmarking of one of these new methods, Swift Biosciences Accel-NGS 1S Plus kit, on mock viral communities and aquatic samples [20]. Briefly, viral DNA (10-20 ng) was fragmented with a E220 Covaris instrument with parameters adapted to circular and linear fragments of ssDNA and dsDNA (peak incident power 175 W, duty factor 5 %, 200 cycles per burst, 90 s of treatment time). Fragmented DNA was concentrated by 1.8x AMPure XP purification and used for library preparation using the Accel-NGS 1S Plus DNA Library Kit (Swift Biosciences, Ann Arbor, MI, USA) with slight modifications: after denaturation and extension step, only one 1.2x AMPure XP purification was performed, adaptors were then added, the ligation reaction was cleaned by 1.0x AMPure XP, and the ligated product was amplified by 10 PCR cycles followed by 0.6x AMPure XP purification.

### *5. Library preparation for metatranscriptomic samples*

Different cDNA synthesis protocols were applied to limit off-target sequencing of ribosomal RNA (rRNA) reads. In the case of samples containing a large number of eukaryotic cells (0,2–3 µm, 3–20 µm, >20 µm, >300 µm membrane filters and coral samples), methods including a poly(A)+ mRNA selection step were chosen. Whereas this approach is very efficient in lowering the number of rRNA reads, it does not allow to retrotranscribe mRNAs from bacterial and archaeal species. cDNA synthesis from bacterial, archaeal and virus RNAs (in the 0.2–3 µm fraction) was independently performed by a random priming approach, preceded by a bacterial and archaeal rRNA depletion step. This method allows cDNA synthesis from both eukaryotic and bacterial and archaeal mRNA and organellar transcripts. The quantity of total

RNA extracts was an additional factor, which conditioned the choice of the cDNA synthesis method (Fig. **3**)

### *5.1. Metatranscriptomic and dual transcriptomic library preparation for eukaryotic mRNA*

#### *5.1.1. Samples with high RNA concentration*

The TruSeq Stranded mRNA Sample Prep used for high RNA inputs allows retaining strand information of RNA transcripts (sequence reads occur in the same orientation as antisense RNA). Based on the total RNA quantity available, the RNA input used for this library was 1 µg (when coral samples were predominant) or 400 ng (when planktonic samples were predominant). Briefly, poly(A)+ RNA was selected with oligo(dT) beads, chemically fragmented by divalent cations under high temperature, converted into single-stranded cDNA using random hexamer priming, and followed by second strand synthesis. Double-stranded cDNA was purified by 1,8x AMPure XP clean ups, and 3'-adenylated. TruSeq RNA barcoded adaptors with 6 bases (Illumina, San Diego, CA, USA) or NEXTflex DNA barcoded adaptors with 12 bases (BiOO Scientific Corporation, Austin, TX, USA) were added in order to comply with sequencing requirements (NEXTflex DNA barcoded adaptors with 12 bases allowing a higher multiplexing for sequencing). The dilution of NEXTflex DNA barcoded adaptors was adjusted according to the RNA input. After one 1x AMPure XP clean up, the ligated product was amplified by 15 PCR cycles and purified by 0,8X AMPure XP clean up.

#### *5.1.2. Samples with low RNA concentration*

The NebUII Stranded mRNA kit was used for low RNA inputs, allowing to retain strand information of RNA transcripts. 50-100 ng of total RNA was used for cDNA synthesis using the NEBNext Ultra II Directional RNA Library Prep for Illumina. Briefly, poly(A)+ RNA was selected with oligo(dT) beads, chemically fragmented by divalent cations under high temperature, converted into single-stranded cDNA using random hexamer priming, and amplified for second strand synthesis. Double stranded cDNA was purified by 1,8x SPRIselect (Beckman Coulter Genomics, Danvers, MA, USA) clean ups, end-repaired, and 3'-adenylated. NEBNext Multiplex RNA barcoded adaptors (Illumina, San Diego, CA, USA) were subsequently added. After one 1x AMPure XP clean up, the ligated product was amplified by 15 PCR cycles and purified by 0,8X AMPure XP clean up.

### *5.2. Metatranscriptomic library preparation for bacterial and archaeal mRNA*

The first step of this protocol consists in a bacterial rRNA depletion followed by cDNA synthesis with the SMARTer Stranded RNA-Seq Kit (Clontech/Takara Bio, CA, USA). The latter is based on chemical RNA fragmentation followed by a first cDNA strand synthesis by random priming and SMART template switching technology. Then, single-stranded cDNA is directly amplified with oligonucleotides containing Illumina adaptors and index sequences to obtain a ready-to-sequence library, preserving the coding strand information. Bacterial rRNA depletion was carried out using the Ribo-Zero Magnetic Kit for Bacteria (Epicentre Biotechnologies, Madison, WI, USA). rRNA depletion was performed on varying total RNA inputs, oscillating between undetectable quantities (Qubit measurement) up to 4 µg. Therefore, the Ribo-Zero depletion protocol was modified to suit low RNA input amounts [21]. Except for these modifications, depletion was performed according to the manufacturer instructions. Briefly, depleted RNA was concentrated to 10 µl total volume with the RNA Clean and Concentrator-5 kit (ZymoResearch) following the procedure described for retention of >17 nt RNA fragments. If total RNA input was > or equal to 1 µg, the amount of depleted RNA was determined by Qubit RNA HS Assay quantification and 40 ng or less were used to synthetize cDNA with the SMARTer Stranded RNA-Seq Kit. Otherwise, 7 µl were used for cDNA synthesis. Single stranded cDNA was purified by two rounds of purification with 1x AMPure XP beads. The purified product was amplified by 18 PCR cycles with the SeqAmp DNA polymerase and the Illumina Index Primer set, both provided in the kit. The final library was purified with 1x AMPure XP beads.

### *6. Metabarcoding strategies*

Metabarcoding strategies were performed on DNA from *i*) cnidarian coral tissues: *Pocillopora meandrina*, *Porites lobata* (Anthozoa: Scleractinia), and *Millepora platyphylla* (Hydrozoa: Milleporidae); *ii*) coral-surrounding water; *iii*) mucus and gut tissues of two reef fishes (*Acanthurus triostegus* and *Zanclus cornutus*); *iv*) surface water above the reef and between the studied island systems; and *v)* coral reef sediments. Three markers were targeted: *i*) a hypervariable region (V9) of the 18S rDNA for eukaryotes, *ii*) two hypervariable regions (V4/V5) of the 16S rRNA gene for prokaryotes, and *iii*) the hypervariable intergenic second internal transcribed spacer (ITS2) of the ribosomal gene array for discriminating Symbiodiniaceae [22] [Data citations #4].

#### *6.1. Amplicon generation*

We started the Tara Pacific expedition using the amplicon generation strategy previously adopted for the Tara Oceans expedition [17]. Briefly, this strategy consisted of producing PCR products from one DNA sample using specific primers, and to construct one library per sample using a NextFlex DNA barcode (one barcode per library). However, in order to reduce cost and execution time of library construction and sequencing, we validated and implemented an alternative sample barcoding strategy using BID (Barcode IDentifier) (Fig. **S1**) for Tara Pacific. The idea relies on introducing such BIDs during the PCR step (12 different BIDs were added to the amplification primers (Tables **S2**) allowing to pool 6 to 12 PCR products upstream of the library preparation. Thus, from the pool of PCR products marked by different BIDs, we constructed one single library indexed by a NextFlex DNA barcode. From multiple gene markers we explored the total diversity of bacteria, archaea, and eukaryotes associated with various compartments of the coral holobiont.

- The hypervariable V9 loop of the 18S rRNA gene allows the analysis of *i*) the taxonomic status of each coral host; *ii*) the eukaryotic portion of the coral holobiont; *iii)* the eukaryotic diversity in coral surrounding water and over extensive taxonomic and ecological scales in surface waters, and coral reef sediments. 18SV9 barcodes were obtained with the primer pair 1389F / 1510R [23] [Data citations #5].

- The hypervariable V4 and V5 loops of the 16S rRNA gene allow the analysis of the *i*) bacterial and archaeal diversity in the coral holobiont; *ii*) bacterial and archaeal diversity in coral-surrounding water and over extensive taxonomic and ecological scales in surface waters and coral reef sediments; and *iii*) microbiome of gut and mucus of fishes. The 515F-Y / 926R 16S primers [24] were chosen to target bacteria and archaea, although they can also co-amplify mitochondrial and chloroplastic DNA of eukaryotic cells. Water samples with a predominance of bacterial and archaeal fraction (e.g., S0.2-3µm) or with a non-negligible abundance of bacterial and archaeal fraction among eukaryotes (e.g., S3-20µm, S>20µm, coral reef sediments and gut fish samples), have been directly amplified with this primer set. In contrast, samples with a very low bacterial and archaeal fraction (e.g., S300µm, coral, and fish mucus), needed a protocol adjustment. In this case, we implemented a nested PCR performing a first full-length amplification using the 27F / 1492R 16S universal primer set [25,26] in order to increase the target DNA, and a second amplification using the 515F-Y / 926R primers [Data citations #6].

- Specific primers of Symbiodiniaceae targeting the ITS2 region of the nuclear ribosomal DNA locus (SYM-VAR-5.8S2 / SYM-VAR-REV) allow the analysis of *i*) the ITS2 type profiles forming the coral holobiont, *ii*) the Symbiodiniaceae diversity in the S3-20 µm size fraction of planktonic samples from coral-surrounding water and surface water, and *iii*) in coral reef sediments. The ITS2 PCR protocol used in this study [27,28] delivers improved specificity and sensitivity with apparent minimal sub-genera taxonomic bias [22] across samples from a wide range of environmental sources. We worked as much as possible with the high-fidelity enzyme Finnzyme Phusion of the High-Fidelity PCR Master Mix with GC Buffer (ThermoFisher Scientific, Waltham, MA, USA). Since the amplification of multiple samples repeatedly failed under these conditions, we switched to the enzyme from the QIAGEN Multiplex PCR Kit (Qiagen, Hilden, Germany) which has lower fidelity, but is less sensitive to inhibitors [Data citations #7].

- PCR amplifications of aerosol samples were performed in triplicate with the Bioline MyTaq HS (BIOLINE, Meridian bioscience, USA) (at the Weizmann Institute). PCR amplification was performed in triplicate and PCR products were pooled after amplification and cleaned using AMPure XP beads using a ratio DNA/beads adapted to the length of the amplicon. Amplicon lengths were verified using a high-throughput LabChip GX microfluidic capillary electrophoresis system (Perkin Elmer, Waltham, MA, USA) and quantified with a Fluoroskan instrument. A NC was included in each PCR experiment, as well as a positive control specific to the targeted gene marker. All details of the mixture and amplification conditions used are described in Tables **S3** and **S4**.

### *6.2. Library preparation from amplicon PCRs*

All libraries were prepared using the NEBNext DNA Modules Products and NextFlex DNA barcodes with 100 ng of purified PCR product as input. The sole difference relied on the pooling of BID-PCRs as opposed to that of no-BID PCRs. For the no-BID strategy, 100 ng of each purified PCR product was oriented directly towards a library preparation. For the BID strategy, purified BID PCR products were normalized at 2.5 ng / µl. Then, an equimolar pool of 6 to 12 BID-PCRs was prepared in order to have a total of 100 ng of amplicons in a total volume of 50 µl. The pooled PCR products were end-repaired, A-tailed at the 3'end, and ligated to Illumina-compatible adaptors using the NEBNext DNA Modules and NextFlex DNA barcodes using a Biomek FX Laboratory Automation Workstation liquid handler (Beckman Coulter Genomics, Danvers, MA, USA), able to perform up to 96 reactions in parallel. After two consecutive 1x AMPure XP clean ups, (except for 18SV9 for which only one 1x Ampure was performed), the ligated products were amplified using the Kapa Hifi HotStart NGS library Amplification kit, followed by 1x AMPure XP purification.

## 7. Sequencing and data quality control

### *7.1. Quality control of sequencing libraries*

All manually prepared libraries were first quantified by Qubit dsDNA HS Assay measurement. A size profile analysis was then conducted in an Agilent 2100 Bioanalyzer (Agilent Technologies, Santa Clara, CA, USA) and by qPCR with the KAPA Library Quantification Kit for Illumina Libraries (Kapa Biosystems, Wilmington, MA, USA) on an MXPro instrument (Agilent Technologies, Santa Clara, CA, USA). All libraries prepared using the Biomek FX Laboratory Automation Workstation were quantified first by PicoGreen in 96-well plates. Library profiles were assessed using a high throughput microfluidic capillary electrophoresis LabChip GX system (Perkin Elmer, Waltham, MA, USA) and qPCR with the KAPA Library Quantification Kit for Illumina Libraries on an MXPro instrument.

### *7.2. Illumina sequencing*

Libraries were subjected to Illumina sequencing in order to obtain the desired number of paired-end reads as described in Table **2**. Metabarcoding libraries were characterized by low nucleotide diversity at the beginning of the reads due to the presence of the primer sequences used for amplification. Such, low-diversity libraries can interfere during the identification of the clusters, resulting in a drastic loss of data output. Therefore, loading concentrations of these libraries and PhiX DNA spike-ins were adapted in order to minimize the impacts on the run quality (10% of PhiX for MiSeq sequencing and 20 % for the other instruments). A primary analysis was performed during the sequencing run by the Illumina Real Time Analysis (RTA) software (Code availability **1**). This tool analyzes images and cluster intensities, and removes low-quality data (i.e., filtering of reads that do not reach the thresholds imposed by the Illumina chastity filter). Furthermore, it performs basecalling and calculates a Phred quality score (Q-score), which indicates the probability that a given base is called incorrectly. The bcl2fastq Conversion Software v2.20.0.422 (Code availability **2**) was used to convert raw BCL files generated by RTA to fastq demultiplexed files allowing one mismatch during the index sequence identification. For the metabarcoding data generated with BID libraries, a second demultiplexing step was performed to assign amplicon sequences to the correct samples and remove the 8-bp BID sequence. BID sequences were searched in the two paired-end reads files (READ 1 and 2) with cutadapt v.1.18 (Code availability **3**). An amplicon sequence was attributed to a given BID if it was found in both sequencing files (READ 1 and 2) (only one mismatch was allowed on the 8-bp BID sequence). The amplicon sequencing library protocol produced non oriented fragments. We obtained 50% of sequences oriented in Forward/Reverse sens and 50% of sequences oriented in reverse/Forward sens. A first validation step took place after the end of the sequencing run, where parameters such as throughput, number of clusters, rate of passing filter clusters, and global Q30 were checked. The sequencing run was validated if all the parameters satisfied Illumina specifications (https://emea.support.illumina.com/bulletins/2019/10/does-my-sequencing-run-look-good-.html). In addition, the metrics of sequencing runs were stored in our LIMS.

### *7.3. Data quality control and filtering for metagenomic and metatranscriptomic sequencing data*

After Illumina sequencing, an in-house quality control process was applied to the reads that passed the Illumina quality filters [17] (Fig. **S2**). In the first step, Illumina sequencing adaptors and primer sequences were removed. Next, low-quality nucleotides ($Q < 20$) were discarded from both ends of the reads. The longest sequence without adaptors and low-quality bases was kept. Sequences between the second unknown nucleotide (N) and the end of the read were also trimmed. Reads shorter than 30 nucleotides (after trimming) were discarded. All trimming and removal steps were performed through an in-house-developed software called fastx_clean (Code availability **4,5**). In the last step, we discarded reads that were mapped to the Enterobacteria phage PhiX174 genome (GenBank: NC_001422.1), using bowtie2 v2.2.9 (-L 31 --mp 4 --rdg 6,6 --local --no-unal) [29]. In the case of metatranscriptomic data, remaining rRNA reads were removed using SortMeRNA v2.1 [30] and SILVA databases [31] (Code availability **6**). In addition, quality controls were performed on random subsets of 20,000 reads before (“raw” reads) and after filtering steps (“clean” reads): *i*) duplicate sequence rate was estimated from raw single- and paired-end sequences using fastx_estimate_duplicate (Code availability **7**); *ii*) read size, quality values, undetermined bases positions and base composition were calculated and sequencing adaptors were detected before and after read filtering; *iii*) taxonomic assignment was performed using Centrifuge v1.0.3 [32] and the NCBI non redundant nucleotide database; *iv*) in the case of overlapping paired-end reads, the merging step was performed with fastx_mergepairs (Code availability **8)**. The first 36 nucleotides of READ2 were extracted and aligned with READ1. Merging was performed if the alignment was at least 15 nucleotides long, had less than 4 mismatches, and an identity of at least 90%. For each overlapping position, the nucleotide of higher quality was retained. Quality control was performed on each dataset and corresponding results were stored in our LIMS and visualized using its web interface (Section **2.5**).

### *7.4. Data quality control for metabarcoding sequencing data*

No quality or adaptor trimming was performed on metabarcoding sequencing data (Fig. **S3**). However, a quality check was performed on a random subset of 20,000 raw sequencing reads as described in the previous section, except that the taxonomic assignment was performed with SortMeRNA v2.1 [30] and SILVA databases (v119, [31]) (for 16S and 18S experiments), PR2 database (v4.3.0, [33]) (for 18S experiments) and ITS2 database (options: --best 1 --fastx --blast '1 cigar qcov' --aligned rRNA -other not_rRNA --log -v --otu_map --de_novo_otu --id 0.97 --coverage 0.97). The ITS2 database was composed of ITS2 sequences downloaded from NCBI. This draft taxonomic assignment allowed us to gauge sample quality and avoid sample inversions.

NCs for extraction and PCR were performed during the metabarcoding library preparation. The taxonomic assignment of NCs allowed us to build a database of possible contaminant species that can be present in reagents (Fig. **4**). This database was used to detect highly contaminated samples (Section **2.5.2)**. The NC sequences were adaptor- and quality-trimmed with fastx_clean and the cleaned sequences were merged with usearch v9.2.64 [34] (Code availability **9**) (-fastq_mergepairs *.fastq -fastqout merged.fq -relabel @ options). The merged sequences were then quality-filtered (-fastq_filter merged.fastq -fastq_maxee 1.0 -fastaout merged.fa -relabel options), dereplicated (-derep_fulllength merged.fa -sizeout -relabel Uniq -fastaout merged.uniques.fa options), and clustered (-cluster_otus merged.uniques.fa -minsize 2 -otus otus.fa -relabel Cluster options). Clusters were taxonomically assigned using SortMeRNA (same options as above) and the SILVA databases for 16S and 18S experiments [31], PR2 database for 18S experiments [33], and ITS2 database for ITS2 experiments. The abundance of the clusters were calculated using USEARCH and an Operational Taxonomic Unit (OTU) table was generated (-usearch_global *.fq -db assigned_clusters.fa -strand plus -id 0.97 -log make_otutab.log -otutabout otutab.txt options) (Fig. **S4**). These results were stored in our LIMS and available via its web interface.

## Data Records

Sample provenance and environmental context are available on Zenodo [Data citations #1]. Samples and their metadata were registered in the ENA biosample database [Data citations #2]. All sequencing files were submitted to the European Nucleotide Archive (ENA) at the EMBL European Bioinformatics Institute (EMBL-EBI) under the Tara Pacific Umbrella BioProject PRJEB47249 [Data citations #3].

## *Technical Validation*

### *1. Sample and experiments information management*

An in-house LIMS called NGL (Next Generation LIMS) was developed to answer the need for collecting, associating, and perusing the substantial amount of data associated with each sample. NGL allows linking the metadata of each sample with information added from its collection to the submission of the sequencing files at the EMBL-EBI. It stores the information and allows the user to follow the samples during the processes, to perform some statistics for reporting and potential troubleshooting. NGL is composed of several specialized modules (Fig.

**5**). The first one is called NGL-P (for Project management). The second, NGL-S, was designed for Sample management, i.e., registration upon its arrival and direct linking to metadata. Next, all experiments are registered in the NGL-SQ (SeQuencing) module, which allows the user to store relevant information such as the type of input material, reagents used, sequencing output, and all QC steps performed. A web interface allows the user to interact with the database and fill in the data of multiple samples at the same time. This module is designed to monitor the flow of experiments until the sequencing step. After sequencing, the run information is stored in the NGL-BI (for BIoinformatics) module which in term interacts with the NGS-QC (for Quality Control) pipelines. NGL-BI orchestrates the execution of the bioinformatics tasks through an interaction, via an API REST, between NGL-BI and the workflows. The QC applied to the sequencing data produced values and graphs stored in NGL-BI, and can be visualized through a web interface. This interface allows users to check the quality control results and validate the cleaned sequencing files. For metabarcoding samples, a specific QC pipeline was applied (Sections **7.4**, **2.5.2**). The comparison between samples and negative controls was performed using the NGS-BA (Biological Analysis) pipeline, stored in the NGL-BI section and visualized via the NGL-BI interface. Finally, the NGL-SUB (SUBmission) module performs the submission to the ENA database (EMBL-EBI) of the cleaned sequencing files, which are linked to their associated biosample and metadata.

## *2. Quality control during sample processing*

### *2.1 DNA quality control*

DNA quantification was performed using dsDNA-specific fluorometric quantitation methods. This quantification allowed: *i*) the validation of extracted DNA and *ii*) the choice of protocols using the related decisional trees for metagenomics (Fig. **2**) and metabarcoding (Table **S3**). An extracted DNA was validated when its concentration was >1 ng/µl. Otherwise, a second DNA purification was attempted on a replicate of planktonic sample, or other homogenized suspension aliquots for coral and fish samples.

### *2.2 RNA quality control*

RNA quality was evaluated by capillary electrophoresis on an Agilent Bioanalyzer using the RNA 6,000 Pico LabChip kit. The Total Eukaryotic RNA Assay was selected for internal electropherogram analysis of RNA extracted from coral, fish, protist, and metazoan-enriched filters, whereas the Total Prokaryotic RNA Assay was applied to prokaryote-enriched filters. This software allowed the generation of an RNA Integrity Number (RIN), calculated by comparing rRNA peaks with a specific database (eukaryotic or bacterial and archaeal). RIN is usually used as a score of RNA quality. In many Tara Pacific samples, eukaryotic and prokaryotic species were co-extracted, generating atypical rRNA peak profiles. For this reason, the RIN was not accurate (or even not computable), and did not reflect the quality of the preparations. However, RNA quality was sometimes poor as most Agilent profiles showed rRNA peaks but also variable amounts of small sized RNAs, indicating partial degradation. A visual evaluation of the Bioanalyzer profiles allowed us to complete the information of the RIN in order to classify them according to a color code (Fig. **S5**). Although this classification was not used as a parameter to validate RNA extracts, it was nevertheless able to explain failed library preparation or poor metrics after assembly. RNA quantity was evaluated using the Qubit RNA HS Assay. The measure of RNA yield allowed: *i*) the validation of RNA extracts and *ii*) the choice of protocols using the relating decisional trees for metatranscriptomics (Fig. **3**). An RNA extract was validated when the concentration was >0.5 ng / µl and >1 ng respectively for coral and planktonic samples. Otherwise a second DNA purification was attempted on a replicate of planktonic sample, or other homogenized suspension aliquots for coral and fish samples.

### *2.3 Amplicon quality control*

Regardless of the amplicon generation strategy used (no-BID versus BID), purified PCR products were quantified with a Fluoroskan instrument (except for the 16S full-length PCR) and validated using a high-throughput microfluidic capillary electrophoresis LabChip GX system (Perkin Elmer, Waltham, MA, USA). LabChip profiles allow checking if the PCR product sizes are in accordance with the primer sets used (Table **3**), and if the latter are absent, thus attesting an effective purification. PCR yield is the second parameter allowing the validation of amplicons. Purified PCR products were validated if their concentration was > 2.5 ng / µl for a no-BID strategy and > 1.5 ng / µl for a BID strategy.

### *2.4 Library quality control*

The qualitative and quantitative controls performed on ready-to-sequence libraries were a crucial step for achieving high-quality sequencing data. First, library size profiles obtained via Agilent or LabChip instruments were carefully evaluated. Libraries were validated if their size profile corresponded to the expected, depending on the library construction protocol used (Table **4**). Qubit quantification and a qPCR assay were routinely performed at the end of library preparation (as recommended by Illumina), the later value was retained for library normalization. Indeed, qPCR-based quantification was accurate and led to optimum cluster densities across each lane of the flow cell.

### *2.5 Sequencing quality control*

#### *2.5.1 Validation of metatranscriptomic and metagenomic experiments*

Metadata (e.g., Tara Pacific identifier, taxon, sampling location, sequencing library type etc) and data produced during the QC workflow are stored in our LIMS and accessible through a web interface (NGL-BI). **(**Fig. **S6**). Platform users can easily check QC metrics and validate the corresponding sequencing files. The upper panel displays the number of clean sequences generated for a given sequencing library (Fig. **S6a**). The panels “Read quality (vs Raw)” and “Read quality (vs Cleaned)” display relevant graphs such as the distribution of the Q30 score at each position before and after the cleaning process (Fig**. S6a** and **S6b**). This score should be generally kept above 80%, but a slight decrease can sometimes be observed at the end of the sequence. Base composition along the sequence is also provided (Fig. **S6c**). For metatranscriptomic experiments, the composition of the first bases can be biased as they contain primers used during the RNA retrotranscription. Moreover, sequencing primers and adaptors used during library preparation are detected and displayed using a heatmap (Fig. **S5d**). After cleaning, some statistics (e.g., number of trimmed sequences, number of removed bases) are shown in the “Trimming” panel (Fig. **S6e**). Generally, a high number of rejected reads should be an alert of the poor quality of the sequencing data. For metatranscriptomic samples, the proportion of rRNA reads is an indicator of the efficiency of the rRNA depletion process (Fig. **S6f**). If the proportion of rRNA read is higher than an arbitrary cutoff of 10%, the sequencing file could be invalidated. An estimation of the read duplication rate (Fig. **S6g**) was also calculated on raw paired-end reads. The duplication rate should ideally be < 20%, a higher value indicates that the sample is of low-complexity sample and, in this case, the metagenomics sequencing file could be invalidated. Statistics on the merging process of paired-end reads are also displayed (Fig. **S6h**). A high proportion reflects a library enriched in small fragments, as an example, in metatranscriptomic experiments. Enrichment of small fragments may indicate that the RNA was degraded. The results of the taxonomic assignment are displayed in a specific panel (Fig. **S7**). The number of assigned sequences and their corresponding assignments are reported. A significant rate of unknown sequences can be observed, especially for plankton samples (samples from water filtrations or from sediments).

### *2.5.2 Validation of metabarcoding experiments*

As metabarcoding sequencing reads were not cleaned or trimmed at this step, only the quality control workflow was performed. As with other types of data, the panel, called “Read quality (vs Raw)” is available through the web interface. Base composition along the reads is usually biased by the fact that we amplified a very small and highly conserved region of the genome. The statistics corresponding to the merging step are important especially for metabarcoding experiments. Indeed, a specific length of the fragment is expected depending on the region that was targeted, and merging paired-reads allows us to verify that the length is as expected. For the 16S and 18S experiments, the observed length can differ from the expected one. Indeed, off-target eukaryotic sequences can sometimes be amplified with the 16S primers and vice versa, whereas bacterial and archaeal sequences can sometimes be amplified with the 18S primers. In such situations, the length distribution of the merged sequences has a two-peaks profile, one at the expected length for 16S sequences and the other at the expected length for 18S sequences. Taxonomic assignments are displayed in the same panel as for metagenomics and metatranscriptomics. As described previously, contaminant DNA present in extraction or PCR reagents were screened in each sample. Amplicon sequences from a given sample were compared to the three NC databases (one NC for extraction and two NC for PCR) using SortMeRNA (Fig. **4**). The proportion of potential contaminant sequences was calculated and a report was generated for each combination of sample and negative control. The results were stored in our LIMS and visualizable through its web interface (Fig. **S8**). Contaminant clusters were previously checked manually based on their taxonomic assignment (probable non-contaminant clusters, like marine bacteria or imprecise assignations, were filtered out) and amplicon sequences assigned to curated contaminant clusters were removed from sequencing files. We particularly monitor samples from the 16S amplification of fish mucus, corals, and size fractions > 300 µm, as the proportion of bacterial DNA in those samples was very low. As a consequence, contaminant DNA from the reagents [35,36] are more likely to have been amplified in those samples. Conversely, samples containing a high proportion of bacterial and archaeal DNA (fish gut or 0.2-3 µm fraction sizes for example) are often free from contamination. Finally, the decontaminated amplicon sequences were again taxonomically assigned using SortMeRNA as described in Section **7.4**, which allows us to verify the efficiency of the decontamination process.

**Code Availability**

1. Real Time Analysis software:

https://www.illumina.com/search.html?filter=support&q=RTA%20download&p=1

2. Bcl2fastq Conversion:

https://support.illumina.com/downloads/bcl2fastq-conversion-software-v2-20.html

3. Cutadapt, https://github.com/marcelm/cutadapt/releases/tag/v1.18

4. Fastx_clean software, http://www.genoscope.cns.fr/fastxtend

5. FASTX-Toolkit, http://hannonlab.cshl.edu/fastx_toolkit/index.html

6. SortMeRNA v2.1, https://github.com/biocore/sortmerna

7. fastx_estimate_duplicate software, http://www.genoscope.cns.fr/fastxtend

8. fastx_mergepairs software, http://www.genoscope.cns.fr/fastxtend

9. Usearch, https://www.drive5.com/usearch/

**Data Citations**

## Acknowledgements

Special thanks to the Tara Ocean Foundation, the R/V Tara crew, and the Tara Pacific Expedition Participants. This research was supported by a research grant from Scott Jordan and Gina Valdez, the De Botton for Marine Science, the Yeda-Sela center for Basic research, and the Sustainability and Energy Research Initiative (SAERI). We are keen to thank the commitment of the following institutions for their financial and scientific support that made this unique Tara Pacific Expedition possible: CNRS, PSL, CSM, EPHE, Genoscope/CEA, Inserm, Université Côte d'Azur, ANR, agnès b., UNESCO-IOC, the Veolia Foundation, the Prince Albert II de Monaco Foundation, Région Bretagne, Billerudkorsnas, Amerisource Bergen Company, Lorient Agglomération, Oceans by Disney, L'Oréal, Biotherm, France Collectivités, Fonds Français pour l'Environnement Mondial (FFEM), Etienne Bourgois, and the Tara Ocean Foundation teams. Tara Pacific would not exist without the continuous support of the participating institutes. This work was supported by the Genoscope, the Commissariat à l'Energie Atomique et aux Energies Alternatives (CEA) and France Génomique (ANR-10-INBS-09-08). This is publication number #00 of the Tara Pacific Consortium

## Author Contributions

CB and JP wrote the manuscript. JP, KL, AA, CC, PO, AP and all the Genoscope technical team were involved in the sequencing tasks. CB, SE, FG, PM, CDS, QC, and JMA developed the bioinformatics pipelines. EJ, JG, SG, GS developed the LIMS. CS was in charge all the Genoscope IT infrastructure. JP, CB, CRV, BCCH, GP, MZ developed ITS2, CDIV, SSED amplicon sequencing protocol. MS developed the viral DNA extraction protocol. MF developed the extraction protocol for aerosol samples. PG contributed by his implication in quality analysis of the metabarcoding sequences. SP and DA contributed as Tara Pacific Scientific directors. EB, CM, GB, GI, SR contributed by their implication in the Tara Pacific samples collection. Tara Pacific Coordinators were involved and implemented the Tara Pacific expedition.

**Tara Pacific Consortium coordinators**

Sylvain Agostini (orcid.org/0000-0001-9040-9296) Shimoda Marine Research Center, University of Tsukuba, 5-10-1, Shimoda, Shizuoka, Japan
Denis Allemand (orcid.org/0000-0002-3089-4290) Centre Scientifique de Monaco, 8 Quai Antoine Ier, MC-98000, Principality of Monaco
Bernard Banaigs (orcid.org/0000-0003-3473-4283) PSL Research University: EPHE-UPVD-CNRS, USR 3278 CRIOBE, Université de Perpignan, France
Emilie Boissin (orcid.org/0000-0002-4110-790X) PSL Research University: EPHE-UPVD-CNRS, USR 3278 CRIOBE, Laboratoire d'Excellence CORAIL, Université de Perpignan, 52 Avenue Paul Alduy, 66860 Perpignan Cedex, France
Emmanuel Boss (orcid.org/0000-0002-8334-9595) School of Marine Sciences, University of Maine, Orono, 04469, Maine, USA
Chris Bowler (orcid.org/0000-0003-3835-6187) Institut de Biologie de l'Ecole Normale Supérieure (IBENS), Ecole normale supérieure, CNRS, INSERM, Université PSL, 75005 Paris, France
Colomban de Vargas (orcid.org/0000-0002-6476-6019) Sorbonne Université, CNRS, Station Biologique de Roscoff, AD2M, UMR 7144, ECOMAP 29680 Roscoff, France & Research Federation for the study of Global Ocean Systems Ecology and Evolution, FR2022/ Tara Oceans-GOSEE, 3 rue Michel-Ange, 75016 Paris, France
Eric Douville (orcid.org/0000-0002-6673-1768) Laboratoire des Sciences du Climat et de l'Environnement, LSCE/IPSL, CEA-CNRS-UVSQ, Université Paris-Saclay, F-91191 Gif-sur-Yvette, France
Michel Flores (orcid.org/0000-0003-3609-286X) Weizmann Institute of Science, Department of Earth and Planetary Sciences, 76100 Rehovot, Israel
Didier Forcioli (orcid.org/0000-0002-5505-0932) Université Côte d'Azur, CNRS, INSERM, IRCAN, Medical School, Nice, France and Department of Medical Genetics, CHU of Nice, France
Paola Furla (orcid.org/0000-0001-9899-942X) Université Côte d'Azur, CNRS, INSERM, IRCAN, Medical School, Nice, France and Department of Medical Genetics, CHU of Nice, France
Pierre Galand (orcid.org/0000-0002-2238-3247) Sorbonne Université, CNRS, Laboratoire d'Ecogéochimie des Environnements Benthiques (LECOB), Observatoire Océanologique de Banyuls, 66650 Banyuls sur mer, France
Eric Gilson (orcid.org/0000-0001-5738-6723) Université Côte d'Azur, CNRS, Inserm, IRCAN, France
Fabien Lombard (orcid.org/0000-0002-8626-8782) Sorbonne Université, Institut de la Mer de

Villefranche sur mer, Laboratoire d'Océanographie de Villefranche, F-06230 Villefranche-sur-Mer, France
Stéphane Pesant (orcid.org/0000-0002-4936-5209) European Molecular Biology Laboratory, European Bioinformatics Institute, Wellcome Genome Campus, Hinxton, Cambridge CB10 1SD, UK
Serge Planes (orcid.org/0000-0002-5689-5371) PSL Research University: EPHE-UPVD-CNRS, USR 3278 CRIOBE, Laboratoire d'Excellence CORAIL, Université de Perpignan, 52 Avenue Paul Alduy, 66860 Perpignan Cedex, France
Stéphanie Reynaud (orcid.org/0000-0001-9975-6075) Centre Scientifique de Monaco, 8 Quai Antoine Ier, MC-98000, Principality of Monaco
Shinichi Sunagawa (orcid.org/0000-0003-3065-0314) Department of Biology, Institute of Microbiology and Swiss Institute of Bioinformatics, Vladimir-Prelog-Weg 4, ETH Zürich, CH-8093 Zürich, Switzerland
Olivier Thomas (orcid.org/0000-0002-5708-1409) Marine Biodiscovery Laboratory, School of Chemistry and Ryan Institute, National University of Ireland, Galway, Ireland
Romain Troublé (ORCID not-available) Fondation Tara Océan, Base Tara, 8 rue de Prague, 75 012 Paris, France
Rebecca Vega Thurber (orcid.org/0000-0003-3516-2061) Oregon State University, Department of Microbiology, 220 Nash Hall, 97331Corvallis OR USA
Christian R. Voolstra (orcid.org/0000-0003-4555-3795) Department of Biology, University of Konstanz, 78457 Konstanz, Germany
Patrick Wincker (orcid.org/0000-0001-7562-3454) Génomique Métabolique, Genoscope, Institut François Jacob, CEA, CNRS, Univ Evry, Université Paris-Saclay, 91057 Evry, France
Didier Zoccola (orcid.org/0000-0002-1524-8098) Centre Scientifique de Monaco, 8 Quai Antoine Ier, MC-98000, Principality of Monaco

### Genoscope Technical Team

Julie Batisse
Odette Beluche
Laurie Bertrand
Chloé Bohers
Isabelle Bordelais
Elodie Brun
Maria Dubois
Corinne Dumont
El Hajji Zineb
Barbara Estrada
Evelyne Ettedgui
Patricia Fernandez
Sonia Garidi
Thomas Guérin
Kevin Gorrichon
Chadia Hamon
Lucille Kientzel
Sandrine Lebled
Chloé Legrain
Patricia Lenoble

Marine Lepretre
Claudine Louesse
Ghislaine Magdelenat
Eric Mahieu
Nathalie Martins
Claire Milani
Céline Orvain
Sophie Oztas
Emilie Payen
Emmanuelle Petit
Guillaume Rio
Dominique Robert
Muriel Ronsin
Benoit Vacherie
Affiliation : [2]Genoscope, Institut François Jacob, Commissariat à l'Energie Atomique (CEA), Université Paris-Saclay, 2 Rue Gaston Crémieux, 91057 Evry, France

## Competing interests

Authors declare no competing interests.

**Figure 1**: Overview of genomic analysis strategies applied on Tara Pacific samples

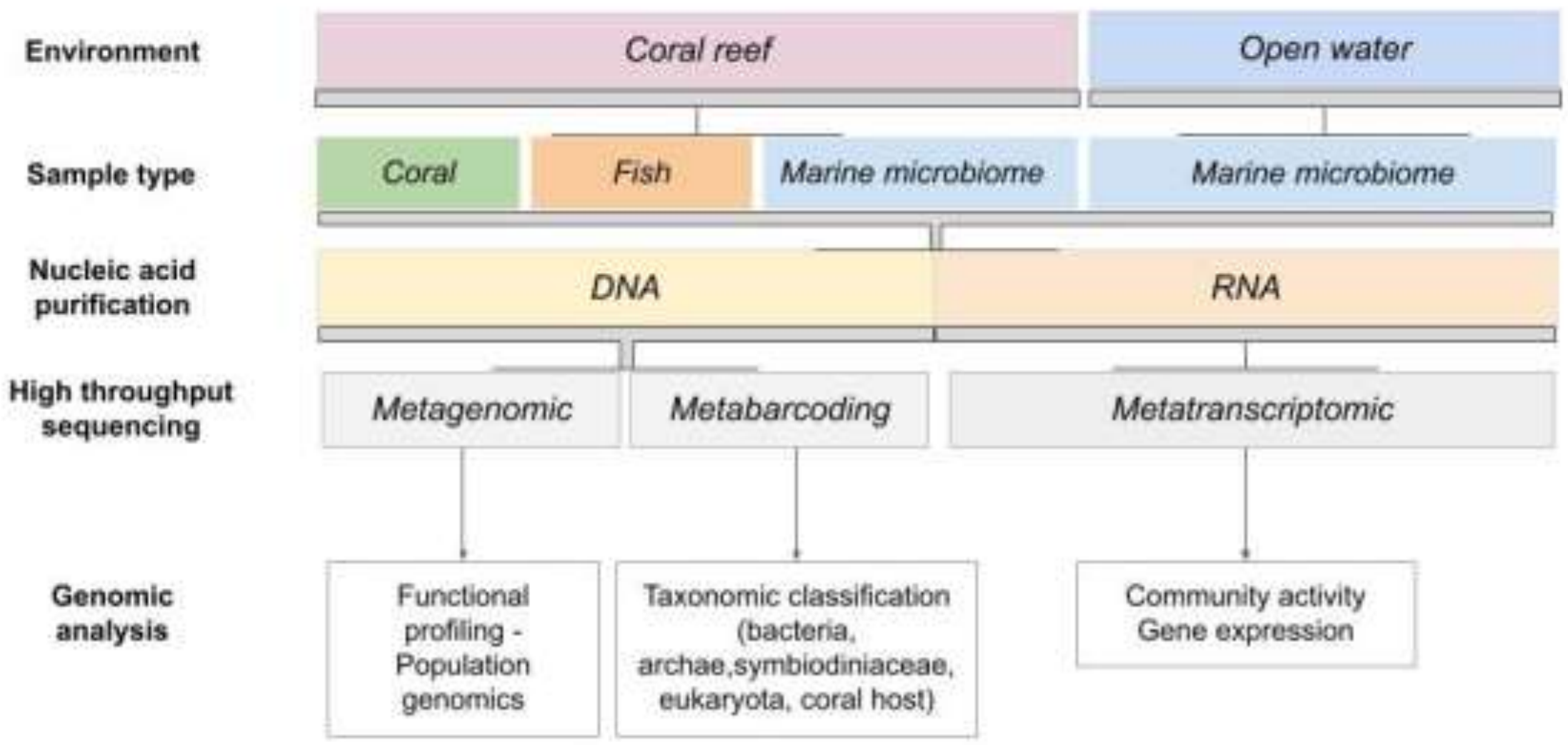

**Figure 2**: Metagenomic library protocol guideline. The choice of the protocol is depending on the amount of the extracted DNA.

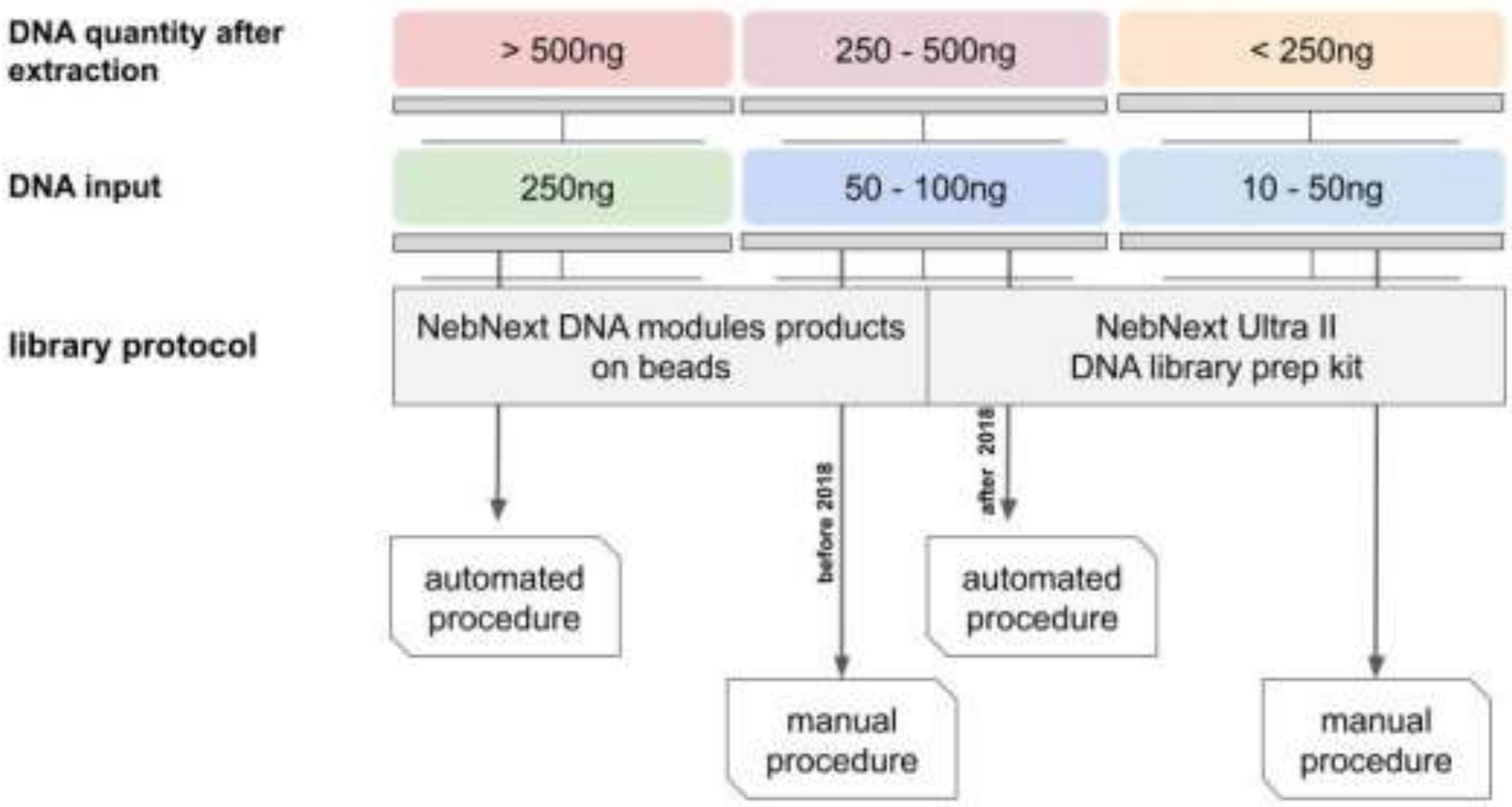

**Figure 3**: Metatranscriptomic library protocol guideline. The choice of the protocol is depending on the amount of the extracted RNA (Dual-T for Dual transcriptomic and MetaT for metatranscriptomic).

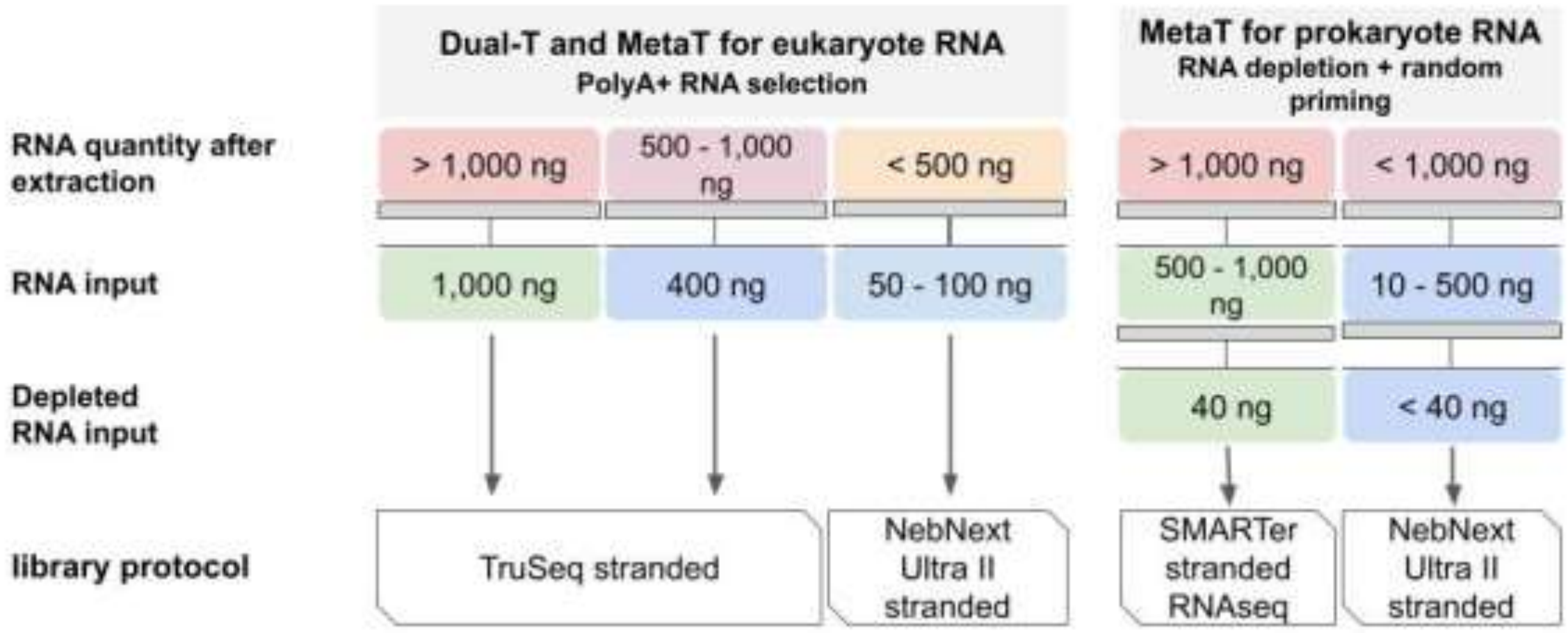

**Figure 4**: Assignment of the Metabarcoding samples against the negative controls database. Samples and negative controls (NC) sequences are clustered separately. NC clusters are assigned against SILVA databases. An OTU table is generated for each NC and a database from the NC clusters is generated. The sample clusters are assigned against these databases. OTU tables are generated and statistics are generated and visible through the NGL-BI web interface.

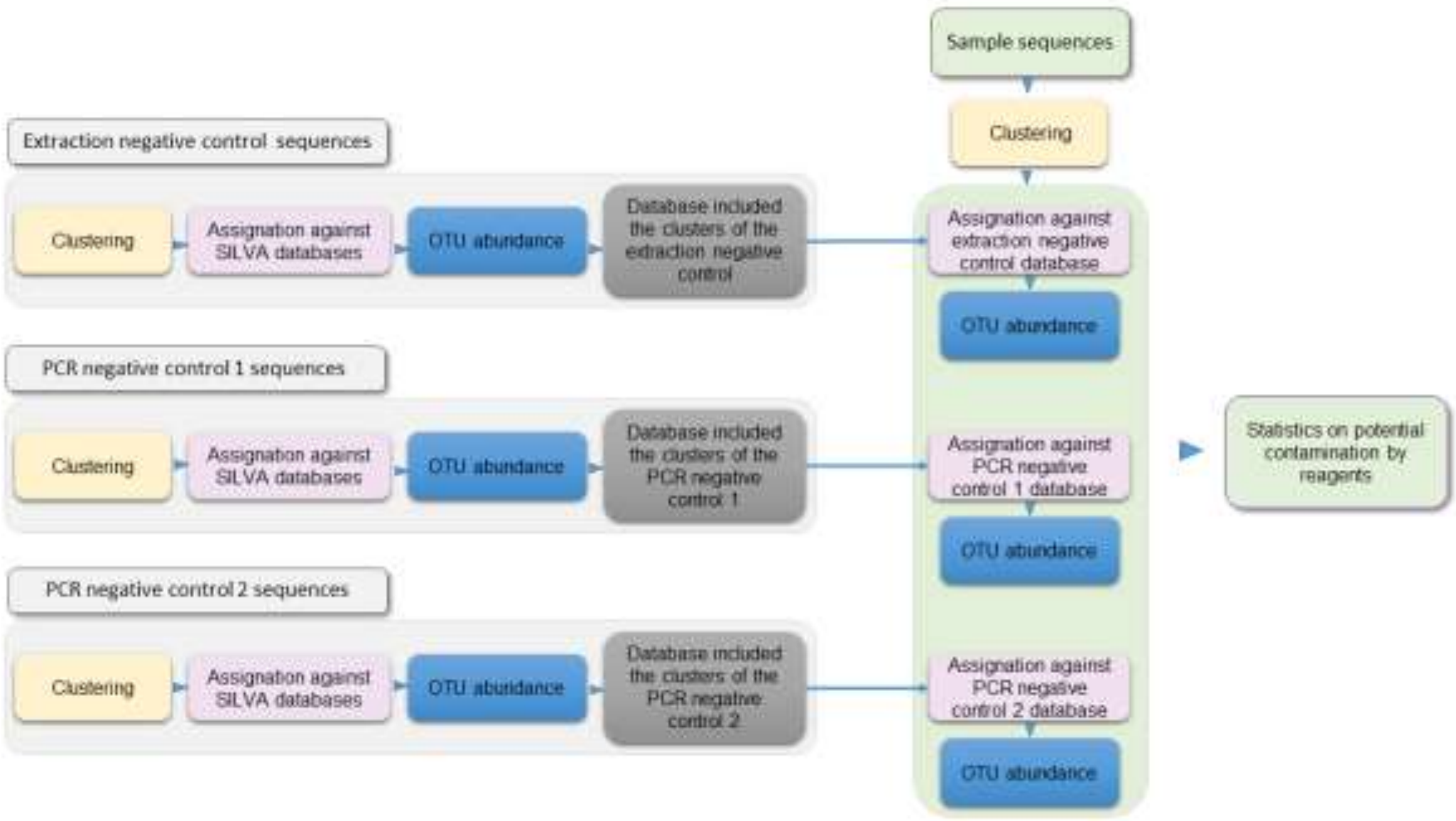

**Figure 5**: NGL (Next Generation Laboratory Information Management System) complete scheme. NGL is composed of different specialized parts: NGL-P for project management, NGL-S for instrument management, NGL-SQ for experiments management, NGL-BI for bioinformatic pipelines management and NGL-SUB for sequencing files submission.

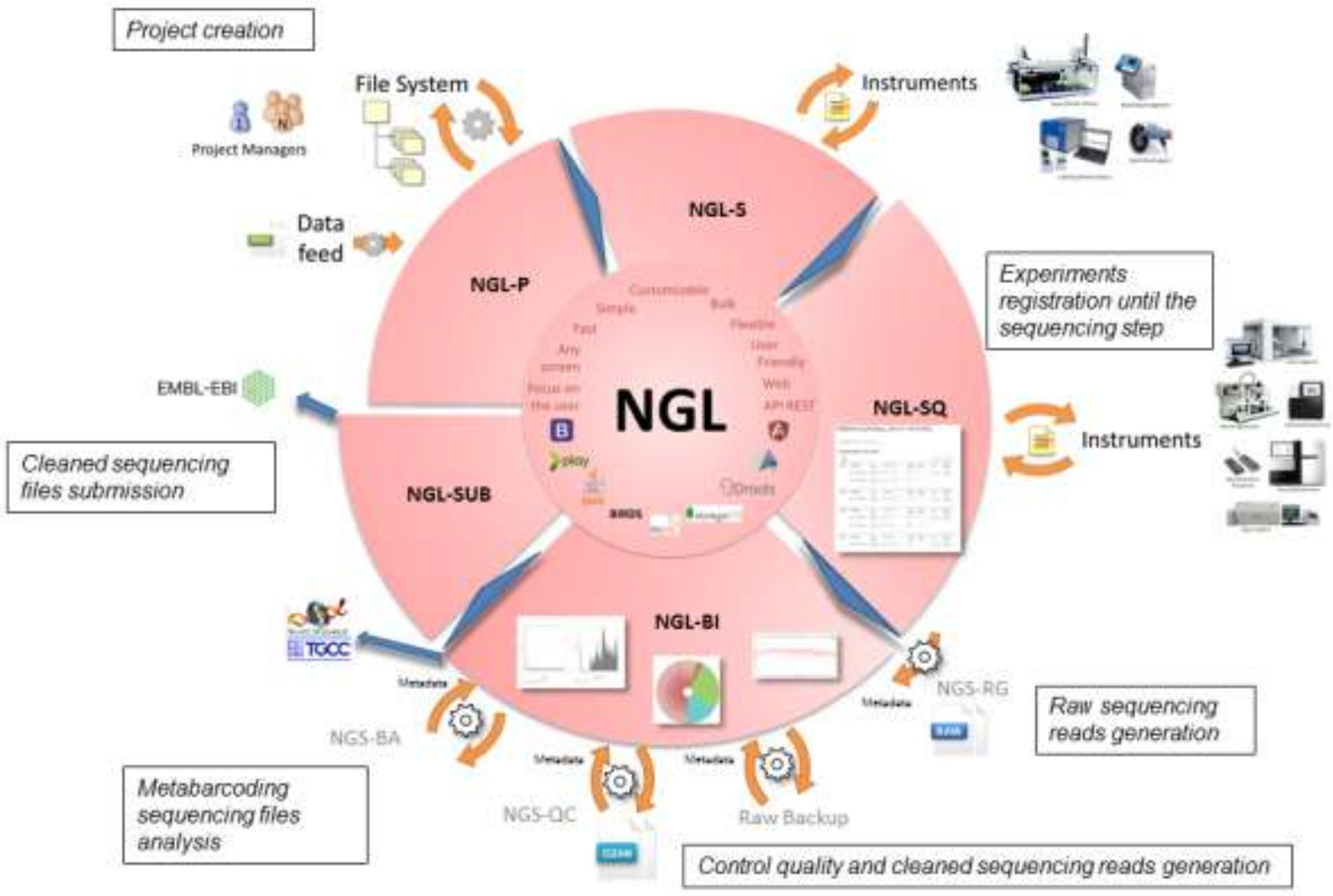

**Table 1**: Overview of the Tara Pacific protocols (EUK for Eukaryote and PROK for Prokaryote).

| SAMPLE MATERIAL | SAMPLE PROTOCOL | Mainly targeted organisms | Sections | Targeted Genomic analysis | | | | | | | |
|---|---|---|---|---|---|---|---|---|---|---|---|
| | | | | Metagenomics | 16S metabarcoding | 18S metabarcoding | ITS2 metabarcoding | Dual-transcriptomics on RNA of EUK | Metatranscriptomics on RNA of EUK | Metatranscriptomics on RNA of PROK | Metaviromics |
| MARINE MICROBIOME | S300 (>300 µm) | Microplankton (EUK) | 2.1.1/2.2.1/3/4.1/5.2/6/7 | X | X | X | | | X | | |
| | S20 (>20 µm) | Microplankton (EUK) | | X | X | X | | | X | | |
| | S320 (3-20 µm) | Nanoplankton (PROK,EUK) | | X | X | X | X | | X | | |
| | S3 (>3 µm) | | | | | | | | | | |
| | S023 (0.2-3 µm) | Picoplankton (PROK,EUK) | 2.1.1/2.2.1/3/4.1/5/6/7 | X | X | X | | | X | X | |
| | S<02 (<0.2 µm) | Viruses | 2.2.5/4.2/7 | | | | | | | | X |
| CORAL HOLOBIONT | CS4L | Millepora, Pocillopora, Porites, associated Symbiodiniaceae and microbiome (PROK,EUK) | 2.1.2/2.2.2/3/4.1/5.1/6/7 | X | X | X | X | X | | | |
| | CDIV | Diverse species of coral, associated Symbiodiniaceae and microbiome (PROK,EUK) | 2.1.2/2.2.3/3/6/7 | | X | X | X | | | | |
| FISH | GT (Gut tissue) | associated microbiome (PROK,EUK) | 2.1.2/2.2.4/3.1/4.1/6/7 | X | X | | | | | | |
| | MUC (Mucus tissue) | associated microbiome (PROK,EUK) | 2.1.2/2.2.4/3.1/6/7 | | X | | | | | | |
| AEROSOLS | AS-ABS | Marine aerosol (PROK,EUK) | 2.2.6/3.1/6/7 | | X | | | | | | |
| SEDIMENT | SSED | Benthic microbiome (PROK,EUK), Symbiodiniaceae | 2.2.7/3.1/6/7 | | X | X | X | | | | |

**Table 2**: Target yield depending on the sequencing strategy/targeted genomics analysis.

| SAMPLE MATERIAL | SAMPLE PROTOCOL | Metagenomics | 16S metabarcoding | 18S metabarcoding | ITS2 metabarcoding | Metatranscriptomics on PROK RNA | Metatranscriptomics on EUK RNA | Metaviromics |
|---|---|---|---|---|---|---|---|---|
| MARINE MICROBIOME | S300 (>300 µm) | 100 M seq<br>2x150 bp<br>HiSeq 4000 (2017)<br>NovaSeq 6000 S4 (2018-2019) | 0.5-1 M seq<br>2x250 bp<br>HiSeq 2500 R (2016-2019)<br>MiSeq (2019)<br>NovaSeq 6000 SP (2019) | 0.3-0.5 M seq<br>2x150 bp<br>HiSeq 2500 R (2016-2018)<br>MiSeq (2018-2019)<br>HiSeq 4000 (2016-2018)<br>NovaSeq 6000 SP (2019) | | | 100 M seq<br>2x150 bp<br>HiSeq 4000 (2016-2018)<br>NovaSeq 6000 S2 (2018)<br>NovaSeq 6000 S4 (2018) | |
| | S20 (>20 µm) | 100 M seq<br>2x150 bp<br>HiSeq 4000 (2017)<br>NovaSeq 6000 S4 (2018-2020) | 0.5-1 M seq<br>2x250 bp<br>HiSeq 2500 R (2016-2019)<br>MiSeq (2019-2020)<br>NovaSeq 6000 SP (2019-2020) | 0.3-0.5 M seq<br>2x150 bp<br>HiSeq 2500 R (2016-2018)<br>MiSeq (2018-2019)<br>HiSeq 4000 (2016-2018)<br>NovaSeq 6000 SP (2019) | | | 100 M seq<br>2x150 bp<br>HiSeq 2500 R (2016)<br>HiSeq 4000 (2016-2018)<br>NovaSeq 6000 S4 (2018) | |
| | S320 (3-20 µm)<br>S3 = (>3 µm) | 100 M seq<br>2x150 bp<br>HiSeq 4000 (2017-2018)<br>NovaSeq 6000 S4 (2018-2020) | 0.5-1 M seq<br>2x250 bp<br>HiSeq 2500 R (2016-2019)<br>MiSeq (2018-2019)<br>NovaSeq 6000 SP (2019-2020) | 0.3-0.5 M seq<br>2x150 bp<br>HiSeq 2500 R (2016-2018)<br>MiSeq (2018-2019)<br>HiSeq 4000 (2016-2018)<br>NovaSeq 6000 SP (2019) | 20 000 seq<br>2x250 bp<br>HiSeq 2500 R (2016)<br>MiSeq (2017-2019) | | 100 M seq<br>2x150 bp<br>HiSeq 4000 (2016-2018)<br>NovaSeq 6000 S2 (2018)<br>NovaSeq 6000 S4 (2018-2020) | |
| | S023 (0.2-3 µm) | 100 M seq<br>2x150 bp<br>HiSeq 4000<br>(2016-2018)<br>NovaSeq 6000<br>S4 (2018-2019) | 0.5-1 M seq<br>2x250 bp<br>HiSeq 2500 R<br>(2016-2019)<br>MiSeq (2019-2020)<br>NovaSeq 6000<br>SP (2019-2020) | 0.3-0.5 M seq<br>2x150 bp<br>HiSeq 2500 R<br>(2016)<br>MiSeq (2018-2019)<br>HiSeq 4000<br>(2016-2018)<br>NovaSeq 6000<br>SP (2019) | | 100 M seq<br>2x150 bp<br>HiSeq 4000<br>(2016-2018)<br>NovaSeq 6000<br>S4 (2018-2020) | 100 M seq<br>2x150 bp<br>HiSeq 4000<br>(2016-2018)<br>NovaSeq 6000<br>S4 (2018-2020) | |
| | S<02 (<0.2 µm) | | | | | | | 100 M seq<br>2x150 bp<br>HiSeq 4000<br>(2017-2018)<br>NovaSeq 6000<br>S1 (2019 |
| CORAL HOLOBIONT | CS4L (*Pocillopora-Porites-Millepora*) | 100 M seq<br>2x150 bp<br>HiSeq 4000<br>(2016-2020)<br>NovaSeq 6000<br>S4 (2020) | 0.5-1 M seq<br>2x250 bp<br>HiSeq 2500 R<br>(2018)<br>MiSeq (2019)<br>NovaSeq 6000<br>SP (2019) | 0.1 M seq<br>2x150 bp<br>HiSeq 2500 R<br>(2017-2019)<br>HiSeq 4000<br>(2106-2018)<br>MiSeq (2107-2019)<br>NovaSeq 6000<br>SP (2019) | 20 000 seq<br>2x250 bp<br>HiSeq 2500 R<br>(2017-2019)<br>MiSeq (2017-2019) | | 40 M seq<br>2x150 bp<br>HiSeq 2500 R<br>(2016)<br>MiSeq (2017)<br>HiSeq 4000<br>(2017-2019)<br>NovaSeq 6000<br>S2 (2018)<br>NovaSeq 6000<br>S4 (2018-2019) | |
| | CDIV (diverse coral species) | | 0.5-1 M seq<br>2x250 bp<br>NovaSeq 6000<br>SP (2019) | 0.1 M seq<br>2x150 bp<br>NovaSeq 6000<br>SP (2019) | 20 000 seq<br>2x250 bp<br>MiSeq (2019) | | | |
| FISH | GT (Gut tissue) | 40 M seq<br>2x150 bp<br>HiSeq 2500 R<br>(2016)<br>HiSeq 4000<br>(2017-2018)<br>NovaSeq 6000<br>S4 (2019-2020) | 0.5-1 M seq<br>2x250 bp<br>HiSeq 2500 R<br>(2017-2018)<br>MiSeq (2018-2019)<br>NovaSeq 6000<br>SP (2019) | | | | | |
| | MUC (Mucus tissue) | | 0.5-1 M seq 2x250 bp HiSeq 2500 R (2018) MiSeq (2018-2019) NovaSeq 6000 SP (2019) | | | | | |
| AEROSOLS | AS-ABS | | 0.5-1 M seq 2x250 bp HiSeq 2500 R (2019) MiSeq (2018-2019) | | | | | |
| SEDIMENT | SSED | | 0.5-1 M seq 2x250 bp HiSeq 2500 R (2019-2020) NovaSeq 6000 SP (2020) | 0.3-0.5 M seq 2x150 bp HiSeq 2500 R (2019) NovaSeq 6000 SP (2019) | 20 000 seq 2x250 bp MiSeq (2019) | | | |

**Table 3**: Metabarcoding amplicon primers and expected PCR product lengths

| Target | Primer set name | Primer set sequences | Expected Length (bp) |
|---|---|---|---|
| 16SFL | 27F/1492R | 5'-AGAGTTTGATCMTGGCTCAG–3′<br>5'-TACGGYTACCTTGTTACGACTT–3′ | 1400 |
| 16SV4V5 | 515F-Y/926R | 5'-GTGYCAGCMGCCGCGGTA A-3'<br>5'-CCGYCAATTYMTTTRAGTTT-3' | 411-600 |
| ITS2 | SYM_VAR_5.8S2/SYM_VAR_REV | 5’-GAATTGCAGAACTCCGTGAACC-3'<br>5’-CGGGTTCWCTTGTYTGACTTCATGC 3’ | 234-266 |
| 18SV9 | 1389F/1510R | 5'-TT GTA CAC ACC GCC C-3'<br>5'-CCTTCYGCAGGTTCACCTAC-3' | 150-200 |

Table 4: Expected sequencing library size distribution (BioAnalyzer profiles)

| Library type | Library protocol | Expected size (bp) | Size Profil example |
|---|---|---|---|
| Metabarcoding 16SV4V5 | NebNext DNA-seq | 520-600/720-1000 | 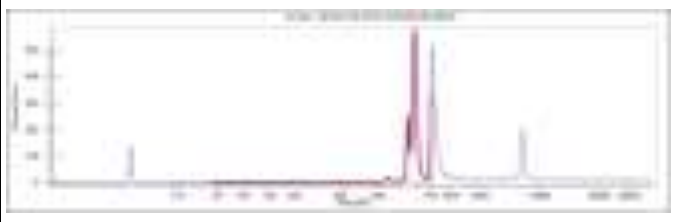 |
| Metabarcoding ITS2 | NebNext DNA-seq | 420-500 | 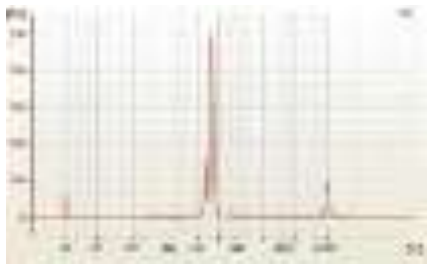 |
| Metabarcoding 18SV9 | NebNext DNA-seq | 270-330 | 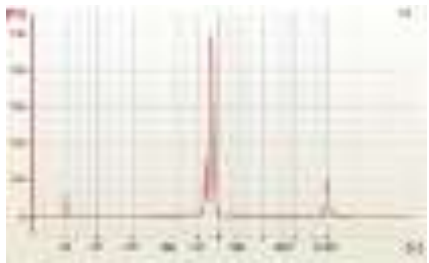 |
| Metagenomic | NebNext Ultra II DNA-seq | 300-800 | 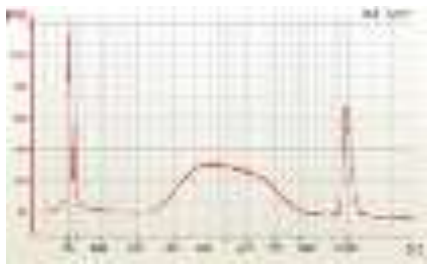 |
| Metagenomic | NebNext DNA-seq | 300-800 | 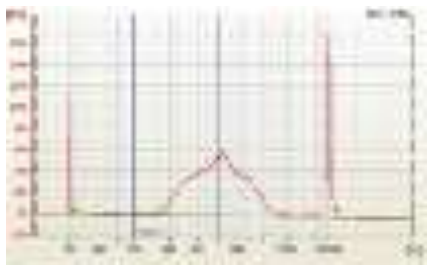 |
| Dual-Transcriptomic Metatranscriptomic | TruSeq Stranded mRNA | 200-600 | 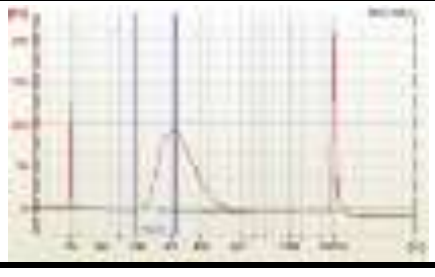 |
| Dual-Transcriptomic Metatranscriptomic | NebNext Ultra II RNA-seq oligo dT | 300-800 | 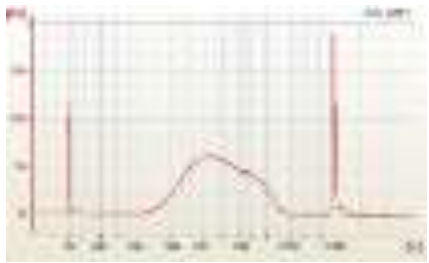 |
| Metatranscriptomic | SMARTer Stranded RNA | 200-600 | 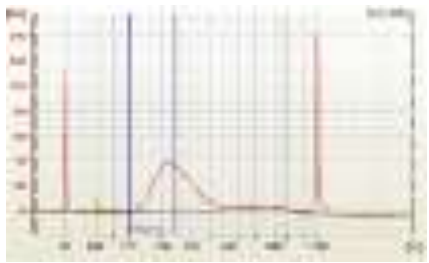 |